\documentclass[prl,twocolumn, superscriptaddress]{revtex4}

\usepackage{dcolumn}
\usepackage{color}
\usepackage{graphicx}

\begin{document}

\title{{\bf Structure of Extremely Nanosized
 and Confined In-O Species in Ordered Porous Materials}}

\author{J.M. Ramallo-L\'opez}
\affiliation{Departamento de F\'{\i}sica, FCE
, Universidad Nacional de La Plata, C.C. N$^\circ$67, 1900 La
Plata, Argentina}
\author{M. Renter\'{\i}a}
\affiliation{Departamento de F\'{\i}sica, FCE
, Universidad Nacional de La Plata, C.C. N$^\circ$67, 1900
La Plata, Argentina} \affiliation{LURE,
 Universit\'e
Paris-Sud,B\^at. 209 A, B.P. 34, 91898 Orsay Cedex, France}
\author{E.E. Mir\'o}
\affiliation{
INCAPE (FIQ, UNL, CONICET), Santiago del Estero 2829, 3000 Santa
Fe, Argentina}
\author{F.G. Requejo}
\affiliation{Departamento de F\'{\i}sica, FCE
, Universidad Nacional de La Plata, C.C. N$^\circ$67, 1900 La
Plata, Argentina}
 \affiliation{
Lawrence Berkeley National Laboratory, One Cyclotron Road-Mailstop
66-200, Berkeley CA 94720, USA}
\author{A. Traverse}
\affiliation{LURE,
 Universit\'e
Paris-Sud,B\^at. 209 A, B.P. 34, 91898 Orsay Cedex, France}

\date{\today}

\begin{abstract}
Perturbed-angular correlation, x-ray absorption, and small-angle
x-ray scattering spectroscopies were suitably combined to
elucidate the local structure of highly diluted and dispersed
InO$_x$ species confined in porous of ZSM5 zeolite. This novel
approach allow us to determined the structure of extremely
nanosized In-O species exchanged inside the 10-atom-ring
channel of the zeolite, and to quantify the amount of In$_2$O$_3$
crystallites deposited onto the external zeolite  surface.

\end{abstract}


\maketitle

The study of structural, magnetic and electronic properties of
nanostructured, subnanostructured and, in the other extreme,
highly diluted monoionic species in solids deserves increasing
attention not only from a fundamental point of view but also for
their technological applications \cite{Li}. In one extreme,
highly-dispersed metallic exchanged-atoms in an infinite variety
of compounds present, e.g., very important catalytic properties
\cite{Greegor}, which are not clearly correlated with the active
species responsible for them since their physical entities are
often unknown. The design of new catalysts with improved activity,
selectivity and stability requires the complete knowledge of the
local environment of the active centers and its correlation with
the desired reaction. The complete characterization of this kind
of structures is at present a challenging problem not only in
catalysis but also in many fields of physics.

An additional difficulty is
the confinement of the diluted species inside porous materials. In
general, experimental techniques based on energetic probes that
strongly interact with the materials, may destroy them. On the
other side, low energetic probes are unuseful since they  have to
pass through the walls around the ``hidden'' locations of the
confined clusters or atoms and their kinetic energy is attenuated.

 The importance of the extended X-ray
absorption fine structure (EXAFS) spectroscopy to study structural
properties in crystalline solids \cite{Greegor}, nanoclusters
\cite{Bazin} and highly dispersed ionic species in catalysis
\cite{Via} has been long acknowledged. EXAFS constitutes a
powerful ``atom selective'' technique to extract direct
information about type, number and distances of neighbors of the
absorber atom \cite{Teo, Groot}. However, in EXAFS analysis, if
more than one species has the same type of bonds (same element and
similar bond-lengths), the information from each one would be
almost impossible to extract directly, unless some additional
information is known.

 The perturbed-angular-correlation (PAC) technique requires
  a suitable probe-atom (native or foreign) to be in the system under
study, being in this sense also ''atom selective''. PAC enables
the precise determination - at the probe site - of the
electric-field gradient tensor (EFG), which is extremely sensitive
to the anisotropy of the  electronic density near the nucleus,
which in turn reflects the probe-neighboring coordination. PAC has
been intensively applied to many fields in science during the past
two decades with success \cite{Lerf} and, very recently, to the
development of accurate {\it ab initio} calculations of the EFG at
diluted impurity sites in crystalline solids \cite{Lany}. However,
structural information is not easily obtained if we are dealing
with highly dispersed, disordered, and coexistent
multiple-species.

In this letter, we will show that the combination of gamma- and
X-ray-based techniques like PAC, EXAFS, and small-angle X-ray
scattering (SAXS)
can be  a novel and powerful
experimental tool that can be applied to the
emerging field of structure determination
of extremely nanosized and
confined species, like In-species at exchange sites in ZSM5
zeolites.

 The crystalline zeolite framework, an aluminum-silicate structure
crossed by channels and cavities, is a challenging laboratory to
study the exchange and deposition of confined particles in
internal surfaces. It has been shown \cite{Ogura} that In
exchanges at Al acid sites of the zeolite and, due to its specific
surface, 90 \% of these sites are inside the channels. In/ZSM5 is
known as a promising catalyst in one of the major topics in
environmental catalysis, the selective catalytic reduction (SCR)
of nitric oxides by hydrocarbons \cite{Ogura, Miro}. The nature of
the active sites for SCR with methane (SCRM) in In/zeolite
catalyst has been extensively studied  but only little qualitative
information about them has been obtained\cite{Armor}. Ogura et al.
\cite{Ogura} showed that the active site in In/ZSM5 for the SCRM
are intrapore In species, and suggested that they are (InO)$^{+}$
oxoions coordinated at the Z$^-$ exchange sites of the zeolite.
Very recently, Schmidt {\it et al.}\cite{Schmidt} reported an
exhaustive spectroscopic study of the structure of In-species (not
necessarily active) using several preparation methods to
synthesize In/ZSM5. Since they did not succeed to obtain a sample
with an isolated active species they have not attempted to extract
a particular In-O distance among several oxygen neighbors to
verify the existence of the In-oxo site. In a previous work we
studied in detail the several In species that can arise using
different preparation methods and activation treatments, leading
to highly active catalysts \cite{Miro}. By means of the PAC
technique we characterized this system and quantified the amount
of In active sites. However, little information about the
structure of this species could be obtained because of the high
distribution of its hyperfine frequency and the lack of accurate
EFG calculations. In this work we have therefore selected the
preparation route that yields to the largest amount of the active
species and with the rest of In atoms forming bulk In$_{2}$O$_{3}$
at the external surface of the zeolite.

NH$_4$-ZSM5 with a Si/Al ratio of 26.4 obtained as in
Ref.\onlinecite{Miro} was used as starting material. Indium was
incorporated to the zeolite by the conventional wet impregnation
method, stirring an aqueous solution of InCl$_3$ (added in an
amount as to obtain a sample with 4 wt$\%$ of In) and NH$_4$-ZSM5
at 80$^{o}$C until all water was evaporated, followed by drying in
a stove at 120$^o$C for 12 h. After this, the solid was pretreated
in a dried oxygen atmosphere heating up to 500$^o$C at 5$^o$C/min,
and holding the final temperature for 12 h. Afterwards, the sample
was calcined for 2 h in oxygen at 750$^o$C. In order to perform
the PAC experiments the probe $^{111}$In was introduced by adding
traces of $^{111}$InCl$_3$ to the non-radioactive InCl$_3$
solution. $^{111}$In decays by electron-capture to excited states
of $^{111}$Cd, which reaches the ground state mainly through a
$\gamma$-$\gamma$ cascade with an intermediate nuclear state of
quadrupole moment Q that interacts with the EFG existing at the
probe site. This EFG, which perturbs the $\gamma$-$\gamma$ angular
correlation, can be determined by measuring the time modulation of
the $\gamma$-$\gamma$ coincidence rate.
 A four BaF$_{2}$-detector
fast-fast coincidence system in a coplanar 90$^o$ arrangement was
used to measure the $\gamma$-$\gamma$ coincidence spectra, which
are combined to form the anisotropy ratio R(t). This R(t) can be
expressed as the sum of cosine functions of three interaction
frequencies from which the strength, V$_{ZZ}$, and symmetry,
$\eta=(V_{XX}-V_{YY})/V_{ZZ}$ , of the diagonalized and traceless
EFG tensor can be determined. If the probes are located at
different sites, their relative concentration f$_i$ can also be
determined model independently. A detailed description of the
technique can be found elsewhere \cite{Frauenfelder}.
\begin{table}[!]
\caption{\label{PAC} Fitted  hyperfine parameters values that
characterize the interactions observed in the PAC spectrum of Fig.
1.}
\begin{ruledtabular}
\begin{tabular}{ccccc}
     Site &   $f$ (\%)   &    $\omega_Q$ (Mrad/s) &    $\eta$   & $\delta$ (\%) \\
\hline
 I$_1$    &  41(5)      &  19.0(2)       &  0.72(1)     &  1.4(2)        \\
I$_2$    &   13(2)      &   25.0(1)      &  0.14(2)     & 1.7(4)
\\
I$_3$    &  46(2)      &  32.2(5)        & 0.33(3)     & 9.1(1.0)
\\
\end{tabular}
\end{ruledtabular}
\end{table}
\begin{figure}[!]
\includegraphics*[bb=29 445 539 632, viewport=-0.05cm 0.1cm 21cm 6.6cm, scale=.475]{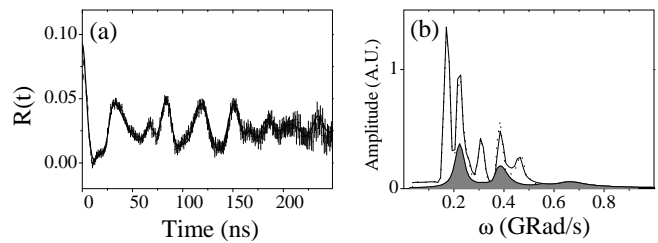}
\caption{\label{pac}(a) PAC spectrum of In($^{111}$In)/ZSM5
measured at 500$^o$C in air. Solid line is the least-squares fit
of theoretical R(t) function to the data. (b) Fourier transforms
of the R(t) spectra (dashed line), of the R(t) fit (solid line),
and the contribution of I$_3$ (shaded curve).}
\end{figure}
Figure 1 shows the PAC spectrum for the In($^{111}$In)/ZSM5 sample
taken at 500$^o$C in air, and its corresponding Fourier transform.
It should be noted that no after-effects due to the
electron-capture decay of $^{111}$In are present, as expected at
this temperature \cite{Habenicht}. In Table I, the fitted
hyperfine parameters f, $\omega_Q=eQV_{ZZ}/40\hbar$, $\eta$ and
$\delta$ (that accounts for a Lorentzian EFG distribution around a
mean $\omega_Q$ value originated from a distribution of very
similar neighborhoods of the probe for a certain site) are shown.
The three hyperfine interactions indicate that In occupies three
different sites. The first two (I$_1$ and I$_2$) correspond to
$^{111}$In in the two inequivalent sites of In$_{2}$O$_{3}$
present in the expected 3:1 population ratio \cite{Habenicht}, and
the third one (I$_3$) with a concentration of 46\% corresponds to
the catalytically active species, since it has been shown that
In$_{2}$O$_{3}$ is not active for the reaction of interest
\cite{Miro}. The low $\delta$ values of I$_1$ and I$_2$ agree well
with the fact that crystallites of very small dimension are really
a minority in the sample.
The slight deviation from axial symmetry of site D has been
already reported for crystalline In$_{2}$O$_{3}$\cite{Bibiloni}.

The In K-edge (27940 eV) X-ray absorption coefficients were
measured at the D42 beamline (XAS13 station) of the DCI
synchrotron at LURE. The spectra were recorded with 4 eV steps at
room temperature in air using a two-crystals Ge400 monochromator
and Kr gas in the ionization chambers. The EXAFS data were
extracted from the measured absorption spectra by standard methods
\cite{Cook}. The spectra were Fourier-transformed in order to
obtain the radial distribution functions around In atoms in the
samples. This function gives a view of the atomic distribution
around the absorber, with peaks at distances where neighboring
shells are located. The height of the peaks is related to the type
and coordination number (CN) of neighbors. The study of small
particles by EXAFS could lead to average coordination numbers
(ACN) smaller than those expected for the bulk
 because of the contribution of atoms in the
surface. The structural and thermal disorder, and a distribution
of slightly different surroundings for a certain site (accounted
for by the Debye-Waller factor, $\sigma^2$) could also reduce the
height of the peaks.
\begin{figure}[!]
\includegraphics*[bb=25 252 555 542, viewport=-1cm 0cm 21cm 10.2cm, scale=0.4]{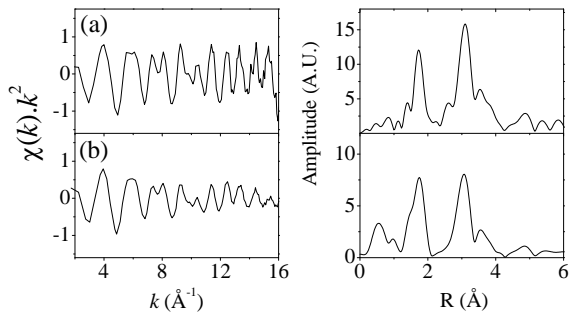}
\caption{\label{EXAFS} In K-edge EXAFS spectra (right) and their
Fourier transforms (left) of  (a) In$_2$O$_3$; (b) In/ZSM5
catalyst.}
\end{figure}
Figure 2 shows the K-edge EXAFS spectra of In$_2$O$_3$ and In/ZSM5
as well as their Fourier transforms. At first sight, both spectra
differ only in
the smaller amplitude of the second one. As there is a mixture of
indium species in the catalyst - In$_2$O$_3$ and In-oxo active
species - and both of them have In-O bonds with overlapping
bond-length, it is not possible to fit directly the EXAFS spectra
to obtain a deconvoluted information of one of them. In contrast,
if doing so we would obtain an average of the In CN of both
species weighted by the relative abundance of each species in the
catalyst. Even though it should be possible to propose a
two-shells fitting of the spectrum with modified CN's, this
procedure will not lead to a unique solution.
 At this
point, the relative fractions of each species obtained from
$^{111}$In-PAC measurements in a model-independent way could be
used to isolate the active species by subtracting the spectrum of
In$_2$O$_3$ weighted with its relative fraction, a method similar
to the difference technique used to isolate a minor component from
the major components in an EXAFS spectrum \cite{Teo}.
Nevertheless, there is one thing that must be considered, i.e.,
the size of the In$_2$O$_3$ crystallites. The $\eta$ parameter of
the symmetric site D observed in the PAC experiments suggested
that small crystallites  of In$_2$O$_3$ could be present.
\begin{figure}[!]
\includegraphics*[bb= 10 20 266 225, viewport=0.3cm -0.1cm 10cm 7.4cm, scale=0.5]{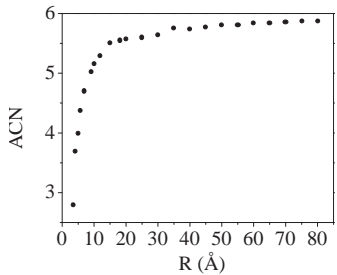}
\caption{\label{calc} ACN of In at the symmetric site in
In$_2$O$_3$.
}
\end{figure}
Figure 3 shows the ACN of In at the symmetric site of
neutral spherical particles of In$_2$O$_3$ for different particle
sizes, obtained averaging the CN of each In atom in the particle.
As can be seen, the ACN grows rapidly for small particles and
approaches the CN of the bulk (CN=6) for radii greater than ca. 20
{\AA}. This implies that, if the In$_2$O$_3$ particles present in
our sample are very small (with radii of the order of 5 to 10
{\AA}), the subtraction of the In$_2$O$_3$ bulk spectrum to the
In/ZSM5 one would lead to an overestimation of the In CN and hence
to an underestimation of the CN of the active species. On the
contrary, if the sesquioxide particles are bigger than 20 {\AA} in
radius, their ACN would be larger than 5.5 and it could be
correct, in a first approximation, to subtract the In$_2$O$_3$
bulk spectrum to obtain that of the other species present in the
catalyst.

Hence, the SAXS technique was used to determine the volume
distribution of indium oxide particles in the In/ZSM5 catalysts.
SAXS experiments were performed at the SAS beamline
\cite{Kellerman} at the National Synchrotron Light Laboratory
(LNLS-Campinas, Brazil). The cluster volume distribution function
D$_V$(R) \cite{Glatter} of the
In-O$_x$ particles is shown in Fig.4.
\begin{figure}[t]
\includegraphics*[bb= 10 18 332 245, viewport=-1cm 0cm 13cm 8cm, scale=0.58]{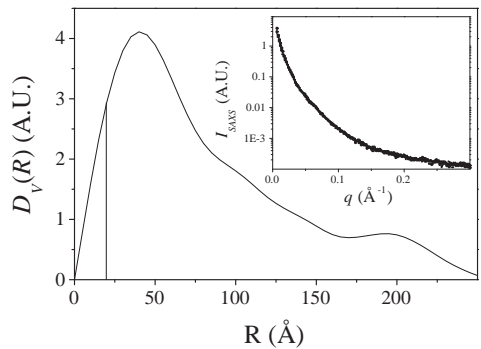}
\caption{\label{distribution}Cluster volume distribution of
In$_2$O$_3$ crystallites in In/ZSM5. The vertical line corresponds
to R=20 {\AA}. The SAXS profile of the In$_2$O$_3$ particles is
shown in the inset.
}
\end{figure}
 This function was
determined from the SAXS intensity profile (Fig. 4, inset) using
the GNOM package \cite{Semenyuk}. This profile was obtained as
usual \cite{Benedetti} by subtracting the ZSM5 normalized SAXS
spectrum to that of the In/ZSM5 catalysts to get rid of the
scattering contribution coming from the porous ZSM5 support.
Figure 4 shows that there is a broad distribution of In$_2$O$_3$
particle sizes in the catalysts and there is a maximum of D$_V$(R)
at R=50 {\AA} with a long tail to larger radii. Although there are
small particles with R $<$ 20 {\AA}, its relative fraction is very
small. Partial integration of the curve shows that 93\% of the
volume corresponds to particles with radius bigger than 20 {\AA}.
\begin{table}[t]
\caption{\label{exafs} EXAFS results of the In K-edge
difference-spectrum of Fig. 5. $\Delta$E$_0$ is the
inner-potential correction.\cite{Teo}}
\begin{ruledtabular}
\begin{tabular}{ccccc}
     Shells &   CN   &    Bond-Length ({\AA}) &    $\sigma^2$({\AA}$^2$)   & $\Delta$E$_0$ (eV) \\
\hline
 In-O    &  0.9(2)      &  2.10(1)       &  0.003(8)     &  2.2(3)        \\
In-O    &   2.1(2)      &   2.20(1)      &  0.011(8)     & 2.2(3)
\\
In-Al    &  1.1(2)      &  2.59(1)        & 0.046(8)     & 4.7(3)
\\
\end{tabular}
\end{ruledtabular}
\end{table}
For particles bigger than 20 {\AA} in radius, the average
coordination number is close to that of the bulk and the
difference can be included within the experimental error for the
CN (ca. 10\%). Under these circumstances, it is valid to subtract
the EXAFS spectrum of bulk In$_2$O$_3$ to the total signal
obtained for the catalyst to obtain the EXAFS spectrum of the
I$_3$ site.
\begin{figure}[!]
\includegraphics*[bb= 102 501 404 737, viewport=-2cm 0cm 13cm 8.7cm, scale=0.572]{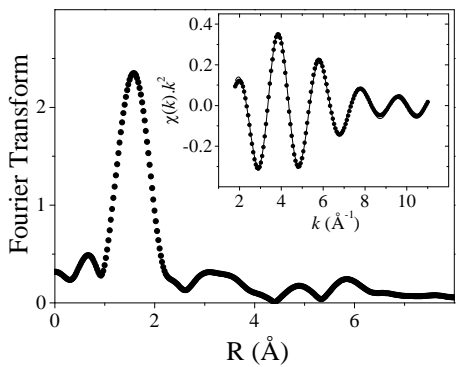}
\caption{\label{fourier}Amplitude  of the Fourier transform of the
difference spectrum corresponding to In species at exchange sites
(circles). In the inset, back-transform of the peak in the range
0.9-2.6 Å (dots) and its EXAFS fit (solid line).}
\end{figure}
Figure 5 shows the Fourier transform of the difference spectrum
obtained subtracting the In$_2$O$_3$ spectrum weighted with the
percentage (54\%) of the oxide in the catalysts - determined by
PAC - to the In/ZSM5 EXAFS spectrum. The inset shows the
backtransform of the peak in the range 0.9-2.6 {\AA} and the
fitted EXAFS functions obtained using phase and amplitudes
generated by the FEFF code \cite{Zabinsky}. The parameters
obtained are shown in Table II. Two oxygen shells and one
aluminium shell were found in the fitting procedure. One oxygen
atom is located at 2.10 {\AA} while 2 oxygen atoms are at 2.20
{\AA}. These last two oxygen atoms would belong to the zeolite
structure and would be bound to an aluminum atom forming an acid
site. The distance of the aluminum atom to the In atom is 2.59
{\AA}. The closer oxygen atom is forming the (InO)$^+$ species
proposed as the active site and the shorter distance is consistent
with a double bond. It is seen that the Debye-Waller factor of the
second and third shells are large. This behavior can be understood
because of the different configurations in which the (InO)$^+$
species can be bound to different acid sites of the zeolite,
leading to slightly different distances In-O (2$^{nd}$ shell) and
In-Al. Similar $\sigma$ values have been found for small Pt
clusters inside zeolite channels \cite{Jentys}. Using theoretical
calculations, Jentys
 {\it et al.} found that Pt particles inside zeolite channels have
Debye-Waller factors two orders of magnitude larger than an
isolated Pt particle. They assigned this effect to an exponential
damping of the EXAFS signal resulting from contributions of the
zeolite lattice atoms. The several anchoring configurations of the
(InO)$^+$ species would also explain the high EFG distribution of
I$_3$ found by PAC (see Table I). Because of the size of the
monometallic species, the active site can only be found in the
10-atom ring (8 {\AA} diam.) of the ZSM5 zeolite as it would not
fit in the 6-atom ring (5 {\AA} diam.).

Consequently, by means of XAFS, TDPAC and SAXS experiments we
determined the local structure of extremely nanozised In-O species
(with the geometry indicated in Table II), responsible for the SCR
of NO$_x$,  located inside the larger ZSM5 zeolite channel.

We wish to dedicate this article to our colleague and friend
Patricia Massolo in the 10th anniversary of her early death. This
research was partially supported by ANPCyT, CONICET, and
Fundaci\'on Antorchas (Argentina), by LURE (CNRS, France), and
LNLS (Brasil) under project SAXS 314/97. Authors express gratitude
to I. Torriani for her assistance at the SAXS beamline.

\newpage


\begin{thebibliography}{}


\bibitem{Li}J. Li {\it et al.},
Science {\bf 299}, 864 (2003); Y. Nishihata {\it et al.},
Nature {\bf 418}, 164 (2002); M. Valden, X. Lai, and D.W. Goodman,
Science {\bf 281}, 1647 (1998).

\bibitem{Greegor}R.B. Greegor, N.E. Pingitore Jr., and  F.
W. Lytle, Science {\bf 275}, 1452 (1997).

\bibitem{Bazin}D. Bazin, Top. Catal. {\bf 18}, 79 (2002).


\bibitem{Via}
W. Li, G.D. Meitzner, R.W. Borry III and E. Iglesia,J. Catal. {\bf
191}, 373 (2000);  V. Schwartz, S.T. Oyama, and J.G. Chen, J.
Phys. Chem. B {\bf 104}, 8800 (2000).

\bibitem{Teo}B.K. Teo, {\it EXAFS: Basic Principles and
Data Analysis} (Springer-Verlag, Munich, 1986).

\bibitem{Groot}F.M.F. de Groot, Topics in Cat. {\bf 10},
179 (2000).

\bibitem{Lerf}A. Lerf and T. Butz, Angew. Chem. Int. Ed. Engl. {\bf 26},
110 (1987); J. Meersschaut {\it et al.}, Phys. Rev. Lett. {\bf
75}, 1638 (1995);
M. Dippel {\it et al.},  Phys. Rev. Lett. {\bf 87}, 95505 (2001);
K. Potzger {\it et al.}, Phys. Rev. Lett. {\bf 88}, 247201 (2002).

\bibitem{Lany}S. Lany {\it et al.}, Phys.
Rev. B {\bf 62}, R2259 (2000); L.A. Errico {\it et al.}, Phys.
Rev. Lett. {\bf 89}, 55503 (2002);
A.T. Motta {\it et al.}, Phys. Rev. B {\bf 65}, 14115 (2001); L.A.
Terrazos {\it et al.}, Solid State Commun. {\bf 121}, 525 (2002).

\bibitem{Ogura}M. Ogura, M. Hayashi, and E. Kikuchi, Catal. Today {\bf 42}, 159
(1998).

\bibitem{Miro}E.E. Mir\'o {\it et al.}, J. Catal. {\bf 188}, 375 (1999).

\bibitem{Armor}E. Kikuchi {\it et al.}, J. Catal  {\bf 161}, 465 (1996); X. Zhou {\it et al.}, J. Mol.
Cat. A: Chem. {\bf 122}, 125 (1997); M. Ogura  {\it et al.}, Cat.
Today {\bf 42}, 159 (1998).

\bibitem{Schmidt}C. Schmidt {\it et al.}, J. Phys. Chem. B {\bf 106}, 4085 (2002).

\bibitem{Frauenfelder}H. Frauenfelder and
R. Steffen, in $\alpha$-, $\beta$-, and $\gamma$-Ray Spectroscopy,
edited by K. Siegbahn (North-Holland, Amsterdam), Vol. 2, 917
(1968).

\bibitem{Habenicht}S. Habenicht {\it et al.}, Z. Phys. B {\bf 101}, 196 (1996).

\bibitem{Bibiloni}A.G. Bibiloni {\it et al.}, Phys. Rev. B {\bf
29}, 1109 (1984).

\bibitem{Cook}T. Ressler {\it et al.},  J. Phys. Chem. B 103, 6407 (1999).


\bibitem{Kellerman}G. Kellerman {\it et
al.}, J. Appl. Crystallogr. {\bf 30}, 880 (1997).

\bibitem{Glatter}O. Glatter and O. Kratky, {\it Small Angle X-Ray Scattering} (Academic, London,
1982).

\bibitem{Semenyuk}A.V. Semenyuk and D.I. Svergum, J. Appl. Crystallogr. {\bf 24}, 537
(1991).

\bibitem{Benedetti}A. Benedetti, J. Appl. Cryst. {\bf 30}, 647 (1997).

\bibitem{Zabinsky}S.I. Zabinsky {\it et al.}, Phys. Rev. B {\bf
52}, 2995 (1995).

 \bibitem{Jentys}A. Jentys,  L. Simon, and  J.A. Lercher,  J. Phys. Chem. B {\bf 104}, 9411
(2000).

\end{thebibliography}
\end{document}